# A linearization for stable and fast geographically weighted Poisson regression

Daisuke Murakami[1], Narumasa Tsutsumida[2], Takahiro Yoshida[3], Tomoki Nakaya[4], Binbin Lu[5], Paul Harris[6]

[1]*Department of Data Statistical Science, Institute of Statistical Mathematics, Tachikawa, Japan*

[2]*Department of Information and Computer Sciences, Graduate School of Science and Engineering, Saitama University, Saitama, Japan*

[3]*Department of Urban Engineering, School of Engineering, The University of Tokyo, Tokyo, Japan*

[4]*Department of Frontier Science for Advanced Environment, The Graduate School of Environmental Studies, Tohoku University, Sendai, Japan*

[5] *School of Remote Sensing and Information Engineering, Wuhan University, Wuhan, China.*

[6]*Net Zero and Resilient Farming, Rothamsted Research, North Wyke, UK*

Although geographically weighted Poisson regression (GWPR) is a popular regression for spatially indexed count data, its development is relatively limited compared to that found for linear geographically weighted regression (GWR), where many extensions (e.g., multiscale GWR, scalable GWR) have been proposed. The weak development of GWPR can be attributed to the computational cost and identification problem in the underpinning Poisson regression model. This study proposes linearized GWPR (L-GWPR) by introducing a log-linear approximation into the GWPR model to overcome these bottlenecks. Because the L-GWPR model is identical to the Gaussian GWR model, it is free from the identification problem, easily implemented, computationally efficient, and offers similar potential for extension. Specifically, L-GWPR does not require a double-loop algorithm, which makes GWPR slow for large samples. Furthermore, we extended L-GWPR by introducing ridge regularization to enhance its stability (regularized L-GWPR). The results of the Monte Carlo experiments confirmed that regularized L-GWPR estimates local coefficients accurately and computationally efficiently. Finally, we compared GWPR and regularized L-GWPR through a crime analysis in Tokyo.

**1: Introduction**

Geographically weighted Poisson regression (GWPR; Nakaya et al., 2005a) was developed as an extension of geographically weighted regression (GWR; Brunsdon et al., 1998) for count data. GWPR performs Poisson regression (PR) considering local samples within a bandwidth distance to investigate the existence of spatially varying relationships between explanatory variables and explained count variables. GWPR has been widely used in regional studies. For example, Nakaya et al. (2005a) employed GWPR to investigate the spatially varying influence of the proportion of professional and technical workers, the unemployment rate, and other covariates on the number of working-age deaths for disease mapping. Hadayeghi et al. (2010) and Pirdavani et al. (2014) used GWPR to analyze the number of vehicle collisions. Other applications include environmental analysis (e.g., Chen et al., 2021) and criminology (e.g., Vilalta and Fondevila, 2019). Despite its popularity, GWPR has inherent issues in terms of (i) computational cost, (ii) stability, and (iii) expandability, which this study seeks to address.

Regarding (i) computation, GWPR can be slow for large samples because it must conduct the iteratively reweighted least squares (IRLS) method together with optimizing the bandwidth parameter. Regarding (ii) stability, GWPR can be unstable for the two properties. First, (ii-1) Poisson regression (PR) is known to suffer from an identification problem for certain data configurations, typically for small samples with many zero counts (Santos-Silva and Tenreyro, 2010; Correia et al., 2019). This problem can make GWPR unstable because this localized PR only considers nearby samples under the bandwidth. Second, (ii-2) local coefficients estimated from GWR models, including GWPR, are prone to be collinear (Wheeler and Tiefelsdorf, 2005), and the resulting coefficient estimates often take extremely large and/or small values (Farber and Páez, 2007; Cho et al., 2009). Regularization is a reasonable remedy for stably estimating GWR/GWPR. LeSage (2004), Wheeler (2007, 2009), and Bárcena et al. (2014), among others, have imposed regularization for the Gaussian GWR. However, such regularization has never been employed for GWPR, probably because of its increase of computational burden.

Regarding (iii) expandability, GWPR has been extended to accommodate over-dispersion (see Section 2.2) by replacing the Poisson model with a negative binomial model (Da Silva and Rodrigues, 2014) or a zero-inflated Poisson model (Kalogirou, 2016). However, the extension of GWPR is limited relative to the linear GWR model. The linear GWR model has been extended to regularized GWR, geographically and temporally weighted regression (Fotheringham et al., 2015), fast/scalable/high-performance GWR (Li et al., 2019; Murakami et al., 2021; Lu et al., 2022), multiscale GWR (Fotheringham et al., 2017) that assumes different spatial scales for each varying coefficient, and mixed/semiparametric GWR (Mei et al., 2004; Nakaya et al., 2009) that assumes spatial variation in some coefficients while others are assumed to be constant. These extended models are widely used for Gaussian data. However, except for a mixed/semiparametric GWPR (Nakaya et al., 2005a; 2009), many of the aforementioned GWR extensions have not yet been transferred to GWPR, which may in part be due to (i) the computational cost and (ii) an identification problem that makes the estimation unstable, especially for small samples (see Section 2.2).

Thus, it is crucial to make GWPR faster, more stable, and easier to extend to apply GWPR to various real-world problems, including those requiring large and complex local modeling. Log-linear approximation is a simple way to improve computational efficiency and extendibility. The linear approximation has not received much attention for approximating PR, mainly because of the lack of accuracy. However, Murakami and Matsui (2022) recently developed a linear approximation that is free from the Poisson identification problem, estimating the PR more accurately than the conventional IRLS-based estimation. Given this recent advance, this study develops a linearized GWPR (L-GWPR) by introducing linear approximation to help overcome issues (i) – (iii). The resulting model has the same form as the basic linear GWR model. Therefore, it is faster, easier to extend, and more stable because it is free from the Poisson identification problem. The remainder of this paper is organized as follows. Section 2 introduces the GWPR, and Section 3 describes the development of L-GWRP and its variants. Section 4 presents Monte Carlo experiments to examine the estimation accuracy and computational efficiency of the proposed approach. Section 5 empirically compares GWPR and L-GWPR through crime analysis in Tokyo. Finally, Section 6 concludes the study.

**2: Geographically weighted Poisson regression (GWPR)**

*2.1: Conventional GWPR*

Consider a study area comprising *N* zones. The GWPR describes the explained count variable $y_i$ observed in the *i*-th zone using the following PR model:

$$y_i \sim Poisson(\lambda_i), \qquad \lambda_i = o_i \exp(\mathbf{x'}_i \boldsymbol{\beta}_i), \qquad (1)$$

where $\lambda_i = E[y_i]$, $o_i$ is an offset variable, and " ′ " denotes the matrix transpose. $\mathbf{x}_i = [1, x_{i,1}, \dots, x_{i,K}]'$ is an explanatory variable vector, and $\boldsymbol{\beta}_i = [\beta_{i,0}, \beta_{i,1}, \dots, \beta_{i,K}]'$ is a coefficient vector in the *i*-th zone. The coefficient $\beta_{i,k}$ is assumed to vary spatially to capture local heterogeneity.

The maximum likelihood (ML) estimator of $\boldsymbol{\beta}_i$ is expressed as follows:

$$\widehat{\boldsymbol{\beta}}_i = (\mathbf{X}'\widehat{\mathbf{A}}\mathbf{W}_i(b)\mathbf{X})^{-1}\mathbf{X}'\widehat{\mathbf{A}}\mathbf{W}_i(b)\hat{\mathbf{z}}_i, \qquad (2)$$

where $\mathbf{X} = [\mathbf{x}_i, \ldots, \mathbf{x}_i]'$, $\hat{\mathbf{z}}_i = [\hat{z}_{1(i)}, \ldots, \hat{z}_{N(i)}]$ with $\hat{z}_{j(i)} = \mathbf{x}'_j \widehat{\boldsymbol{\beta}}_i + (y_j - \hat{\lambda}_j)/\hat{\lambda}_j$ and $\hat{\lambda}_j = o_j \exp(\mathbf{x}'_j \widehat{\boldsymbol{\beta}}_i)$. Suppose that $\text{diag}[\cdot_1, \ldots, \cdot_N]$ denotes a diagonal matrix whose elements are $\{\cdot_1, \ldots, \cdot_N\}$, $\widehat{\mathbf{A}} = \text{diag}[\hat{\lambda}_i, \ldots, \hat{\lambda}_N]$ and $\mathbf{W}_i(b) = \text{diag}[w_{i,1}(b), \ldots, w_{i,N}(b)]$. $w_{i,j}(b)$ is the local weight for zone *j*. Given $w_{i,j}(b)$, $\widehat{\boldsymbol{\beta}}_i$ is estimated by an IRLS method alternately iterating the estimation of $\widehat{\boldsymbol{\beta}}_i$ and updating $\hat{\lambda}_j$ until convergence.

In practice, the local weight $w_{i,j}(b)$ is parameterized using a distance-decay function. The following Gaussian kernel is widely used:

$$w_{i,j}(b) = \exp\left(-\left(\frac{d_{i,j}}{b}\right)^2\right), \qquad (3)$$

where $d_{i,j}$ is the distance between locations *i* and *j*; we used the Euclidean distance between the geographic centers of the two zones. The bandwidth parameter *b* determines the spatial smoothness of the local coefficients. The coefficients have small-scale map patterns if *b* is small, and large-scale patterns if *b* is large. As *b* increases, the coefficient estimator in Eq. (2) converges asymptotically to the conventional PR estimator.

The bandwidth is optimized by minimizing a cost function such as a cross-validation (CV) score or information criterion (see, e.g., Nakaya et al., 2005a). In the optimization, the IRLS is iterated while varying the *b* value to evaluate the cost function. Thus, GWPR relies on a double-loop algorithm that hinders its application for large samples or extends it to more sophisticated models, such as spatiotemporal and multiscale models, which optimize the bandwidth for each coefficient (see Fotheringham et al., 2017; Leong and Yue, 2017).

## 2.2: Quasi-GWPR

Conventional PR and GWPR are restrictive in what they assume to be equidispersion, which means that the mean of the count equals the variance. For basic PR, a quasi-Poisson likelihood was proposed to address over-/under-dispersion (see Wedderburn, 1974; McCullagh and Nelder, 1989). The quasi-likelihood replaces the score function $S(\lambda_i) = \frac{y_i - \lambda_i}{\lambda_i}$ that comprises the Poisson likelihood $L = \int S(\lambda_i) \, d\mu$ with a quasi-score function $S^{(q)}(\lambda_i) = \frac{y_i - \lambda_i}{\sigma^2 \lambda_i}$ by assuming $E[y_i] = \lambda_i$ and $Var[y_i] = \sigma^2 \lambda_i$. $\sigma^2$ is a dispersion parameter in which $\sigma^2 > 1$ indicates overdispersion, and $\sigma^2 < 1$ indicates under-dispersion. The quasi-likelihood is identical to the usual likelihood, assuming an equidispersion if $\sigma^2 = 1$.

By applying the same derivation to the GWPR, a quasi-GWPR accounting for over or under-dispersion was formulated (see Nakaya, 2001; Nakaya et al., 2005b; Smith et al., 2014). The quasi-likelihood yields $L^{(q)} = \int S^{(q)}(\lambda_i) \, d\mu = \frac{1}{\sigma^2}[y_i \log(\lambda_i) - \lambda_i]$. The corresponding quasi-GWPR model is formulated as follows:

$$y_i \sim quasiPoisson(\lambda_i, \sigma^2), \quad \lambda_i = o_i \exp(\mathbf{x}'_i \boldsymbol{\beta}_i), \tag{4}$$

where $E[y_i] = \lambda_i$ and $Var[y_i] = \sigma^2 \lambda_i$.

The model parameters were estimated by maximizing the quasi-likelihood. The dispersion parameter is estimated as follows:

$$\hat{\sigma}^2 = \frac{1}{N - N_{enp}} \sum_{i=1}^{N} \frac{(y_i - \hat{\lambda}_i)^2}{\hat{\lambda}_i}, \tag{5}$$

where $\hat{\lambda}_i = o_i \exp(\mathbf{x}'_i \hat{\boldsymbol{\beta}}_i)$. $N_{enp} = tr[\mathbf{R}]$ is the effective number of parameters in which $tr[\cdot]$ denotes the trace operator and $\mathbf{R}$ is a matrix whose *i*-th row equals $\mathbf{r}_i = \mathbf{x}'_i (\mathbf{X}'\widehat{\mathbf{A}}\mathbf{W}_i(b)\mathbf{X})^{-1} \mathbf{X}'\widehat{\mathbf{A}}\mathbf{W}_i(b)\hat{\mathbf{z}}_i$ (see footnote 1). Similar to quasi-Poisson regression, the coefficient estimator of the quasi-GWPR model is identical to that of the GWPR model.[1] In contrast, the dispersion parameter influences variance of the coefficient. In particular, given Eq. (5), the variance matrix of $\hat{\boldsymbol{\beta}}_i$ yields:

$$Var[\hat{\boldsymbol{\beta}}_i] = \hat{\sigma}^2 (\mathbf{X}'\widehat{\mathbf{A}}\mathbf{W}_i(\hat{b})\mathbf{X})^{-1}, \tag{6}$$

where $\hat{\sigma}^2 = 1$ in the case of the basic GWPR. Eq. (6) is useful for evaluating the statistical significance of localized coefficient estimates.

## *2.3: Identification problem*

Correia et al. (2019) derived conditions for the absence of an ML estimator for generalized linear models (GLMs). In the case of conventional PR, the ML estimator cannot be obtained when there exist a non-zero coefficient vector $\boldsymbol{\gamma} = [\gamma_1, \dots, \gamma_K]'$ satisfying conditions (7) and (8):

---

[1] In the case of the quasi-GWPR model, $\widehat{\mathbf{A}}$ in Eq. (2) is replaced with $\hat{\sigma}^2 \widehat{\mathbf{A}}$. However, the $\hat{\sigma}^2$ parameter is canceled out as follows: $(\hat{\sigma}^2 \mathbf{X}'\widehat{\mathbf{A}}\mathbf{W}_i(b)\mathbf{X})^{-1}(\hat{\sigma}^2 \mathbf{X}'\widehat{\mathbf{A}}\mathbf{W}_i(b)\hat{\mathbf{z}}_i) = (\mathbf{X}'\widehat{\mathbf{A}}\mathbf{W}_i(b)\mathbf{X})^{-1}\mathbf{X}'\widehat{\mathbf{A}}\mathbf{W}_i(b)\hat{\mathbf{z}}_i = \hat{\boldsymbol{\beta}}_i$ (Eq. 2). The coefficient estimator is identical to the basic GWPR model. Similarly, the effective degrees of freedom $N_{enp}$ for the quasi-GWPR model equals $tr[\widetilde{\mathbf{R}}]$ in which the *i*-th element of $\widetilde{\mathbf{R}}$ equals $\tilde{\mathbf{r}}_i = \mathbf{x}'_i(\hat{\sigma}^2 \mathbf{X}'\widehat{\mathbf{A}}\mathbf{W}_i(b)\mathbf{X})^{-1}(\hat{\sigma}^2 \mathbf{X}'\widehat{\mathbf{A}}\mathbf{W}_i(b)\hat{\mathbf{z}}_i) = \mathbf{x}'_i(\mathbf{X}'\widehat{\mathbf{A}}\mathbf{W}_i(b)\mathbf{X})^{-1}\mathbf{X}'\widehat{\mathbf{A}}\mathbf{W}_i(b)\hat{\mathbf{z}}_i = \mathbf{r}_i$ . $N_{enp}$ is unchanged.

$$\mathbf{x}'_i \boldsymbol{\gamma} = 0 \text{ for all } i \text{ such that } y_i > 0, \tag{7}$$

$$\mathbf{x}'_i \boldsymbol{\gamma} \leq 0 \text{ for all } i \text{ such that } y_i = 0. \tag{8}$$

Eq. (7) is fulfilled if the explanatory variables are perfectly collinear (or rank-deficient) for subsamples with $y_i > 0$ (see Santos-Silva and Tenreyro, 2010). Even if Eq. (7) is satisfied, PR is still identifiable if $\mathbf{x}'_i \boldsymbol{\gamma}$ takes positive values when $y_i = 0$ (Eq. 8). As Eq. (8) is easily fulfilled by assuming a large negative intercept in $\boldsymbol{\gamma}$, estimating the Poisson model while avoiding Eq. (7) is important in practice. However, condition (7) cannot be avoided if the number of samples satisfying $y_i > 0$ is less than the number of explanatory variables because of rank deficiency. In other words, PR is only weakly unidentifiable or identifiable when the number of nonzero counts in the observed explained variable is small. This property renders PR models unstable, particularly for small samples with many zero-count observations.

(Quasi-)GWPR may suffer from the same problem because the estimator maximizes the (quasi-)Poisson likelihood given *b*. However, GWPR models only consider local subsamples to estimate the local property. This property makes the identification problem severe, particularly when *b* is small and the subsample being considered decreases. We introduce a linear approximation to the quasi-GWPR model to overcome Poisson identification and computational cost problems.

## 3: Linearized GWPR (L-GWPR)

### *3.1: Outline*

Figure 1 (a) shows the conventional procedure for (quasi-)GWPR. It relies on a double-loop algorithm that optimizes the bandwidth through one iteration and estimates the coefficients in another IRLS iteration.

We propose the L-GWPR model, which is a linear approximation of the quasi-GWPR model. Following Murakami and Matsui (2022), the parameters are estimated in two steps, as shown in Figure 1 (b). First, we approximate the mean $\lambda_i$ of the quasi-GWPR model using a log-linear model. The approximate mean is used to approximate the quasi-GWPR model. The first step, which maximizes the Gaussian likelihood instead of the Poisson likelihood, is free of the Poisson identification problem. Additionally, our approach replaces IRLS with a single-weighted least squares (WLS) estimation. Our single-loop algorithm is faster than the conventional double-loop algorithm (see Figure 1).

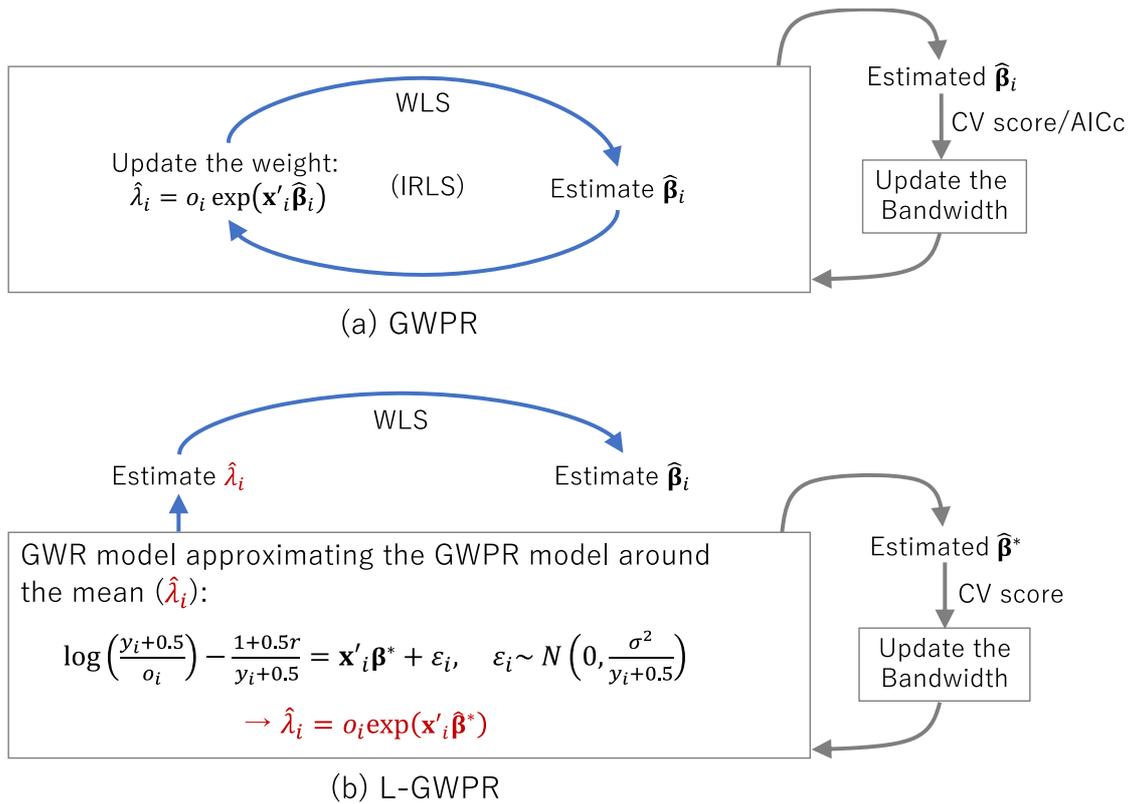

Figure 1. Estimation procedure for (a) (quasi-)GWPR and (b) L-GWPR. The blue arrow represents the procedure for estimating the coefficients, and the gray arrow represents the procedure for optimizing the bandwidth. As illustrated, GWPR has a double-loop routine while L-GWPR does not.

Details of the proposed procedure are described in the following subsections. We first introduce the log-linear approximation of Murakami and Matsui (2022) in Section 3.2 and incorporate it into the quasi-GWPR model in Section 3.3.

### *3.2: Linear approximation for the quasi-Poisson regression*

Consider the quasi-Poisson regression:

$$y_i \sim quasiPoisson(\lambda_i, \sigma^2), \quad \lambda_i = o_i \exp(\mathbf{x}'_i \boldsymbol{\beta}). \tag{9}$$

The coefficient estimator yields:

$$\widehat{\boldsymbol{\beta}} = (\mathbf{X}'\widehat{\mathbf{A}}\mathbf{X})^{-1}\mathbf{X}'\widehat{\mathbf{A}}\hat{\mathbf{z}}. \tag{10}$$

$\hat{\mathbf{z}} = [\hat{z}_1, \dots, \hat{z}_N]$ with $\hat{z}_i = \mathbf{x}'_i \widehat{\boldsymbol{\beta}}^* + (y_i - \hat{\lambda}_i^*)/\hat{\lambda}_i^*$, where $\widehat{\boldsymbol{\beta}}^*$ and $\hat{\lambda}_i^*$ are the estimators of $\boldsymbol{\beta}$ and $\lambda_i$. If the IRLS is used, the estimators yield $\widehat{\boldsymbol{\beta}}^* = \widehat{\boldsymbol{\beta}}$ and $\hat{\lambda}_i^* = o_j \exp(\mathbf{x}'_j \widehat{\boldsymbol{\beta}}_i)$ after convergence. Murakami and Matsui (2022) estimated $\widehat{\boldsymbol{\beta}}^*$ and $\hat{\lambda}_i^*$ using the following log-linear model, approximating the Poisson model in Eq. (11) around the mean $\lambda_i$:

$$z_i^+ = \mathbf{x}'_i \boldsymbol{\beta}^* + \varepsilon_i, \quad \varepsilon_i \sim N\left(0, \frac{\sigma^{*2}}{y_i + 0.5}\right), \tag{11}$$

where $z_i^+ = \log\left(\frac{y_i + 0.5}{o_i}\right) - \frac{1 + 0.5\psi}{y_i + 0.5}$, $\psi$ is the ratio of zero counts in the explained variable, and $\sigma^{*2}$ is the variance parameter. Eq. (11) is derived by equating the mode of the log-linear model with that of Poisson model. The log-linear model in Eq. (11) can be easily estimated using the WLS method.

Their proposed procedure is as follows: (a) estimate $\boldsymbol{\beta}^*$ using the WLS; (b) substitute the estimated $\widehat{\boldsymbol{\beta}}^*$ and $\hat{\lambda}_i^* = o_i \exp(\mathbf{x}'_i \widehat{\boldsymbol{\beta}}^*)$ into Eq. (10) to estimate the PR coefficient, $\widehat{\boldsymbol{\beta}}$. They observed that the estimation accuracy was almost the same as that of conventional PR, or even better for small samples (see Murakami and Matsui, 2022, for further detail).

### 3.3: Linearized GWPR (L-GWPR)

This section proposes the L-GWPR model by introducing a log-linear approximation into the quasi-GWPR model. We can rewrite the GWPR estimator in Eq. (2) as follows:

$$\widehat{\boldsymbol{\beta}}_i = \left(\mathbf{X}'\widehat{\mathbf{A}}\mathbf{W}_i(b)\mathbf{X}\right)^{-1}\mathbf{X}'\widehat{\mathbf{A}}\mathbf{W}_i(b)\widehat{\mathbf{z}}_i, \tag{12}$$

where $\widehat{\mathbf{z}}_i = [\hat{z}_{1(i)}, \dots, \hat{z}_{N(i)}]$ and $\hat{z}_{j(i)} = \mathbf{x}'_j\widehat{\boldsymbol{\beta}}^*_i + (y_j - \hat{\lambda}^*_j)/\hat{\lambda}^*_j$. $\widehat{\boldsymbol{\beta}}^*_i$ and $\hat{\lambda}^*_i$ are the ML estimators, which are $\widehat{\boldsymbol{\beta}}^*_i = \widehat{\boldsymbol{\beta}}_i$ and $\hat{\lambda}^*_i = o_j \exp(\mathbf{x}'_j\widehat{\boldsymbol{\beta}}_i)$ after convergence if the IRLS is used as usual. In Section 3.2, the following linear GWR model approximates the quasi-GWPR model around the mean $\lambda_i$:

$$z_i^+ = \mathbf{x}'_i\boldsymbol{\beta}^*_i + \varepsilon_i, \quad \varepsilon_i \sim N\left(0, \frac{\sigma^{*2}}{y_i + 0.5}\right). \tag{13}$$

Given $b$, the ML estimator of $\boldsymbol{\beta}^*_i$ yields:

$$\widehat{\boldsymbol{\beta}}^*_i = (\mathbf{X}'\mathbf{A}^+\mathbf{W}_i(b)\mathbf{X})^{-1}\mathbf{X}'\mathbf{A}^+\mathbf{W}_i(b)\mathbf{z}^+, \tag{14}$$

where $\mathbf{A}^+ = \text{diag}[(y_1 + 0.5), \dots, (y_N + 0.5)]$ and $\mathbf{z}^+ = [z_1^+, \dots, z_N^+]$.

Based on Nakaya et al. (2005), the variance-covariance matrix of the Poisson estimator $\widehat{\boldsymbol{\beta}}_i$ becomes:

$$Var[\widehat{\boldsymbol{\beta}}_i] = \mathbf{C}(b)\widehat{\mathbf{A}}^{-1}\mathbf{C}(b)', \tag{15}$$

where $\mathbf{C}(b) = \left(\mathbf{X}'\widehat{\mathbf{A}}\mathbf{W}_i(b)\mathbf{X}\right)^{-1}\mathbf{X}'\widehat{\mathbf{A}}\mathbf{W}_i(b)$. Eq. (15) is useful to evaluate the statistical significance of the coefficient.

Several methods exist for optimizing the bandwidth $b$. Based on Eq. (13), minimizing the weighted squared error $\sum_{i=1}^{N}(y_i + 0.5)\left(z_i^+ - \mathbf{x}'_i\widehat{\boldsymbol{\beta}}^*_{-i}\right)^2$ may seem reasonable. However, our preliminary analysis suggests that the resulting $b$ value becomes unstable. Therefore, following Santos and Neves (2008), who derived the theoretical properties of a local PR and used an unweighted squared error, $b$ was optimized by minimizing the following squared error:

$$\sum_{i=1}^{N}\left(z_i^+ - \mathbf{x}'_i\widehat{\boldsymbol{\beta}}^*_{-i}\right)^2, \tag{16}$$

where $\widehat{\boldsymbol{\beta}}^*_{-i}$ is a coefficient vector estimated from a training set; when applying leave-one-out cross-validation (LOOCV), the training set consists of all samples except the $i$-th sample. Minimization of Eq. (16) to maximize the accuracy of $\mathbf{x}'_i\boldsymbol{\beta}^*_i$ is acceptable because $\mathbf{x}'_i\boldsymbol{\beta}^*_i$ approximates the regression term in the GWPR model.

The bandwidth can also be optimized by minimizing the Poisson residual deviance, which is the negative twice of the Poisson likelihood during the LOOCV. The residual deviance of the GWPR is formulated as follows:

$$D\left(\hat{\lambda}^*_{-i}\right) = 2\sum_{i=1}^{N}\left[y_i \log\left(\frac{y_i}{\hat{\lambda}^*_{-i}}\right) - (y_i - \hat{\lambda}^*_{-i})\right], \tag{17}$$

where $\hat{\lambda}^*_{-i} = o_i \exp(\mathbf{x}'_i\widehat{\boldsymbol{\beta}}^*_{-i})$ is the out-of-sample predictor of $y_i$. The LOOCV that minimizes Eq. (17) does not suffer from overfitting and identifies the optimal bandwidth in terms of prediction accuracy because $y_i$ is not used to evaluate the predictor. Such residual deviance (or Poisson loss, which is proportional to the deviance) based on LOOCV is widely accepted in the machine learning literature (see e.g., Ridgeway, 2020).

Once the bandwidth is optimized using LOOCV, minimizing Eq. (16) or Eq. (17), the mean of the GWPR model is approximated as $\hat{\lambda}_i^* = o_i \exp(\mathbf{x}'_i \widehat{\boldsymbol{\beta}}_i^*)$. The obtained $\hat{\lambda}_i^*$ available to the GWPR estimate $\widehat{\boldsymbol{\beta}}_i$ is approximated by substituting $\widehat{\boldsymbol{\beta}}_i^*$ and $\hat{\lambda}_i^*$ into Eq. (12). In short, the L-GWPR approximates the GWPR coefficients as follows ( Figure 1 (b)):

(1) Apply a LOOCV to identify the optimal bandwidth $\hat{b}$,
(2) Substitute $\hat{b}$ into Eq. (14) to estimate $\widehat{\boldsymbol{\beta}}_i^*$, and calculate $\hat{\lambda}_i^* = o_i \exp(\mathbf{x}'_i \widehat{\boldsymbol{\beta}}_i^*)$,
(3) Substitute $\hat{\lambda}_i^*$ and $\widehat{\boldsymbol{\beta}}_i^*$ into Eq. (12) to estimate the local coefficient $\widehat{\boldsymbol{\beta}}_i$.

In contrast to the traditional GWPR iterating IRLS, our CV iterates the WLS. The double-loop algorithm in conventional GWPR is replaced with a single-loop algorithm.

### *3.4: Variants of L-GWPR*

This section proposes two variants of the L-GWPR. Section 3.4.1 explains a regularized L-GWPR, which is useful for improving identifiability. Section 3.4.2 explains a locally linearized GWPR, which is useful for improving the accuracy of the linear approximation.

### *3.4.1: Linearized geographically weighted Poisson ridge regression (L-GWPRR)*

Introducing ridge regularization, as suggested by Wheeler (2007) and Bárcena et al. (2014), is reasonable in the case of the linear GWR model to enhance the stability of GWPR. The regularized L-GWPR, which we refer to as the linearized geographically weighted Poisson ridge regression (L-GWPRR), is implemented by replacing the coefficient estimators in Eqs. (12) and (14), as follows:

$$\widehat{\boldsymbol{\beta}}_i = (\mathbf{X}'\widehat{\mathbf{A}}\mathbf{W}_i(b)\mathbf{X} + \delta\mathbf{I})^{-1}\mathbf{X}'\widehat{\mathbf{A}}\mathbf{W}_i(b)\hat{\mathbf{z}}_i, \qquad (18)$$

$$\widehat{\boldsymbol{\beta}}_i^* = (\mathbf{X}'\mathbf{A}^+\mathbf{W}_i(b)\mathbf{X} + \delta\mathbf{I})^{-1}\mathbf{X}'\mathbf{A}^+\mathbf{W}_i(b)\mathbf{z}^+. \tag{19}$$

where $\delta$ is a ridge parameter. As both $\widehat{\boldsymbol{\beta}}_i$ and $\widehat{\boldsymbol{\beta}}_i^*$ estimate or approximate the same quantity ($\boldsymbol{\beta}$), assuming the same $\delta$ value is acceptable. A large $\delta$ strongly shrinks the estimators toward zero to reduce variance. As $\delta$ approaches zero, regularization becomes weaker, and when $\delta = 0$, the ridge estimator coincides with the conventional estimator (Eqs. 12 and 14). The $\delta$ parameter is optimized together with the bandwidth during LOOCV.

*3.4.2: Locally Linearized GWPR*

Unlike the quasi-Poisson model of Murakami and Matsui (2022), which is derived for all samples, the GWPR considers only local samples approximately within the effective bandwidth (see Cressie 1993), which is the distance at which 95 % of the geographical weight disappears. For the Gaussian kernel, the effective bandwidth is $\sqrt{3}b$. By applying the same derivation as Murakami and Matsui (2022) to the local subsamples, we obtain the same log-linear approximation as $z_{i(loc)}^+ = \log\left(\frac{y_i+0.5}{o_i}\right) - \frac{1+0.5\psi_i}{y_i+0.5}$ instead of $z_i^+$, where $\psi_i$ is the ratio of zero counts in the local sub-samples. Notably, $\sqrt{3}b$ is used only to determine the $\psi_i$ parameter. The bandwidth parameter is interpreted without multiplying $\sqrt{3}$, following the original GWPR. We refer to the locally linearized GWPR using $z_{i(loc)}^+$ instead of $z_i^+$ as L-GWPR_loc, or the ridge term L-GWPRR_loc.

## 4: Monte Carlo experiments

### *4.1: Outline*

Using Monte Carlo experiments, this section compares the estimation accuracy of basic PR, GWPR, L-GWPR, and their variants, which are listed in Table 1. Here, the L-GWPR models using the Poisson deviance cost function in LOOCV are referred to as L-GWPR_dev or L-GWPRR_dev, where only the latter has a ridge term. A Gaussian kernel is used across all the models. R version 4.1.1 (https://cran.r-project.org/) is used for the computations.

Table 1. Models compared

| Name | Model | Cost function in the LOOCV |
|---|---|---|
| PR | Poisson regression | NA |
| GWPR | Conventional GWPR | AICc (see footnote 2) |
| L-GWPR | L-GWPR | Squared error |
| L-GWPR_dev | | Residual deviance |
| L-GWPRR | L-GWPRR | Squared error |
| L-GWPRR_dev | | Residual deviance |
| L-GWPR_loc | L-GWPR_loc | Squared error |
| L-GWPRR_loc | L-GWPR_loc with ridge regularization | |

The code implemented in GWR4 software (https://gwr.maynoothuniversity.ie/gwr4-software/; see Nakaya, 2015) is rewritten into R code to estimate the GWPR. Following the default setting of this software, the bandwidth is optimized through a grid search evaluating the AICc value for each bandwidth ranging between 0.1 and 4.0 with an 0.1 interval.[2] We first attempted to optimize the bandwidth through a golden section search of the minimum AICc value using existing R packages. However, this method failed to find the best bandwidth value in many trials, particularly when the true bandwidth was small, probably because of the Poisson identification problem. Therefore, we believe that grid search is a reasonable choice.

Each model is fitted to the count data generated from the following model:

$$y_i \sim Poisson(\mu_i, \sigma^2), \ \mu_i = \exp(\beta_{i,0} + x_{i,1}\beta_{i,1} + x_{i,2}\beta_{i,2}), \ x_{i,k} \sim N(0,1). \quad (20)$$

where the coefficients are generated from the following spatial moving average processes $\beta_{i,k} = \mu_{k(\beta)} + \sigma_{k(\beta)} \left[ \frac{\sum_{i=1}^{N} g_{i,j}(b) u_{j,k}}{\sum_{i=1}^{N} g_{i,j}(b)} \right]$, where $u_{j,k} \sim N(0,1)$ and $[\cdot]$ is an operator standardizing · with mean zero and variance one.[3] The spatial weight $g_{i,j}(r)$ is given by the $(i, j)$-th element of the spatial proximity matrix, whose $(j, j)$-th element equals $\exp(-d_{i,j}^2/r^2)$, where $r$ is a range parameter that determines the spatial scale of the local coefficient. Spatial coordinates for evaluating distance were generated from two uniform distributions (minimum: –2; maximum: 2). The mean and standard deviation of the three coefficients are assumed to be $\{\mu_{0(\beta)}, \mu_{1(\beta)}, \mu_{2(\beta)}\} = \{\mu_{0(\beta)}, 2, -0.5\}$, where $\mu_{0(\beta)}$ is as explained below, and $\{\sigma_{0(\beta)}, \sigma_{1(\beta)}, \sigma_{2(\beta)}\} = \{1, 2, 1\}$. Therefore, $\beta_{i,1}$ is an influential coefficient explaining larger variations, and $\beta_{i,2}$ is a less influential coefficient.

---

[2] Our preliminary analysis suggested that the grid-search-based AICc-minimization is more stable than the LOOCV, which is another widely-used method for optimizing the GWPR model. We first tried LOOCV for the conventional GWPR, which is implemented in GWmodel. However, the optimization is unstable, and the resulting estimates take singular values or NA values in many trials. Therefore, for the standard GWPR, we used the grid-search-based AICc minimization which is implemented in GWR4. On the other hand, we rely on LOOCV in the case of L-GWPR as it does not have AICc yet. Derivation of AICc will be an important next step.

The parameter values were varied as follows: $\mu_{0(\beta)} \in \{-1, 2\}$, $b \in \{0.5, 1.0, 2.0\}$, and $N \in \{200, 500, 2000\}$. The average ratios of zero counts were 0.568 for $\mu_{0(\beta)} = -1$ and 0.187 for $\mu_{0(\beta)} = 2$ among the 1,000 replicates generated from Eq. (20), assuming $N = 500$ and $b = 1.0$. $\mu_{0(\beta)} = -1$ means a "many zero" scenario, whereas $\mu_{0(\beta)} = 2$ means a "few zero" scenario. Figure 2 shows examples of the simulated coefficients for each $b$ value.

In each case, samples of size $N = 200, 500,$ and $2,000$ were generated 1,000 times, and the estimation accuracy was evaluated using the correlation coefficient (CC), root mean squared error (RMSE), and bias, where the latter two were formulated as follows:

---

[3] Fotheringham and Oshan (2016) have already showed the robustness of GWR against multicollinearity among explanatory variables; hence, multicollinearity is outside the scope of our study.

$$RMSE[\hat{\beta}_{i,k}] = \sqrt{\frac{1}{N}\sum_{i=1}^{N}(\hat{\beta}_{i,k} - \beta_{i,k})^2}, \quad (21)$$

$$bias[\hat{\beta}_{i,k}] = \frac{1}{N}\sum_{i=1}^{N}\hat{\beta}_{i,k} - \frac{1}{N}\sum_{i=1}^{N}\beta_{i,k}. \quad (22)$$

The CC between $\hat{\beta}_{i,k}$ and $\beta_{i,k}$ is used because a large positive CC implies similar map patterns for $\hat{\beta}_{i,k}$ and $\beta_{i,k}$. This similarity is important because the estimated local regression coefficients are typically interpreted through visual inspection (i.e., if the map pattern is similar to the true pattern, erroneous interpretation is less likely).

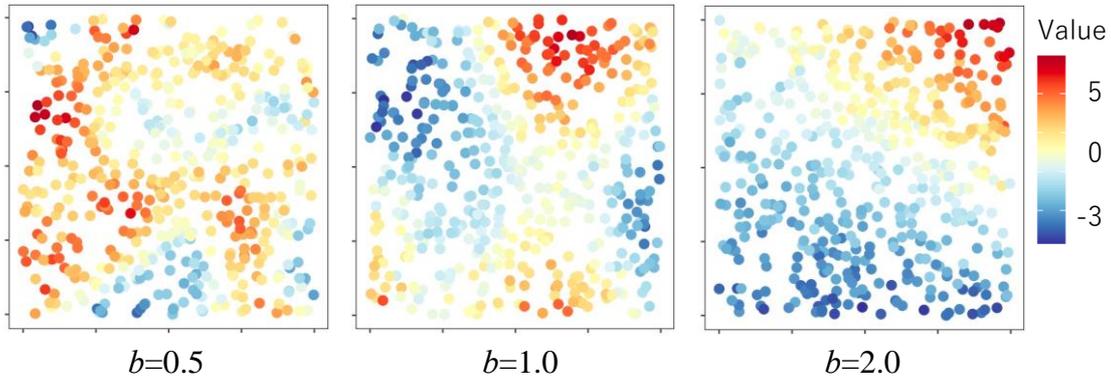

Figure 2. Examples of the generated $\beta_{i,1}$ for each assumed bandwidth value ($N = 500$).

*4.2: Results*

4.2.1: Coefficient estimation accuracy

Figure 3 displays the boxplots of the CC values between the true and estimated coefficients under the "few zero" scenario ($\mu_{0(\beta)} = 2$). The results are similar across the three bandwidths. Conventional GWPR indicated highly positive CCs across all cases, where the correlations strengthened as *N* increased. The accuracy of the GWPR coefficients was good in terms of CCs. Although L-GWPR had a similar accuracy as GWPR in terms of the median CC, the CC values for L-GWPR had a larger variation and tended to be unstable. In contrast, the CC values of L-GWPRR were relatively stable. Furthermore, the CCs of L-GWPRR improved over GWPR when the sample size was 500 or 2,000. The results demonstrate the importance of regularization in estimating the linearized GWPR.

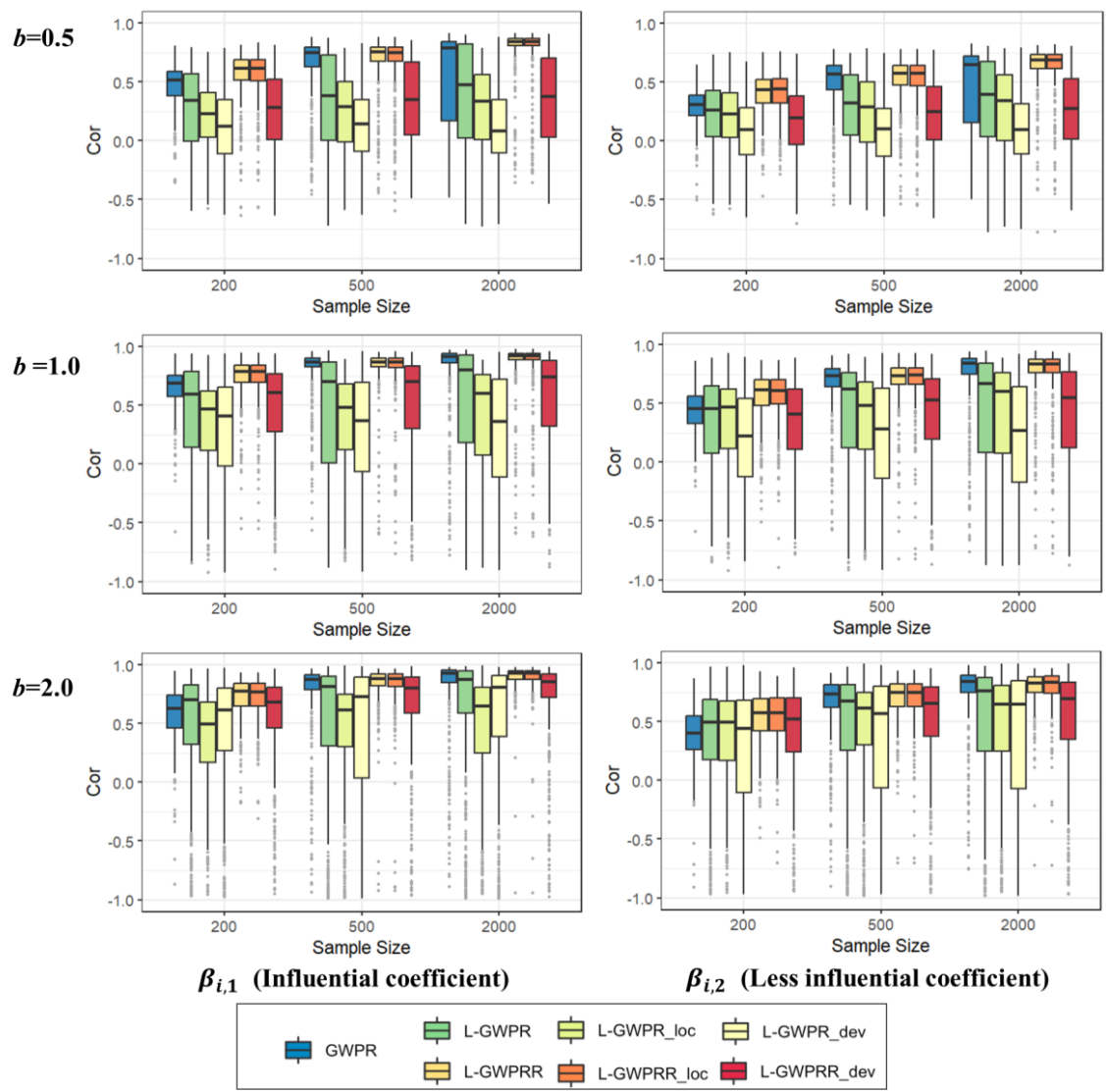

Figure 3. Boxplots of the CCs between $\hat{\beta}_{i,k}$ and $\beta_{i,k}$ in each iteration when $\mu_{0(\beta)} = 2$ (Left: influential coefficient ($\beta_{i,1}$); right: less influential coefficient ($\beta_{i,2}$)).

The CC distribution of L-GWPRR_loc was approximately the same as that of L-GWPRR, indicating that local linearization may not necessarily improve the model accuracy. The CC values of the deviance-based L-GWPR models (L-GWPR_dev and L-GWPRR_dev) were relatively low and unstable, which can be attributed to the Poisson identification problem. In summary, the L-GWPRR models were among the best specifications in terms of CC distributions.

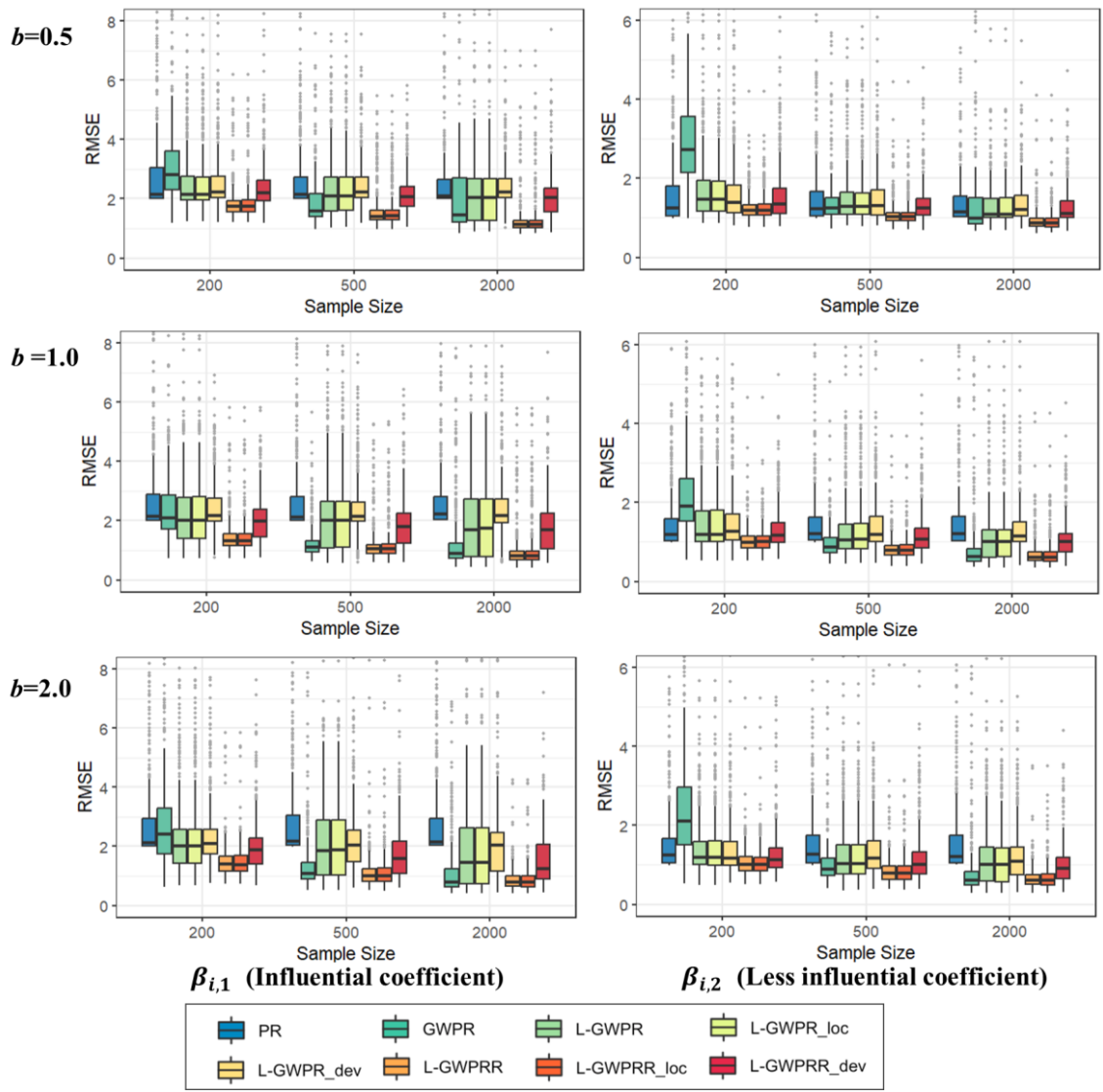

Figure 4. Boxplots of RMSE in each iteration when $\mu_{0(\beta)} = 2$ (Left: influential coefficient ($\beta_{i,1}$); right: less influential coefficient ($\beta_{i,2}$)).

Figure 4 shows boxplots of the RMSE values under this scenario ($\mu_{0(\beta)} = 2$). For $\beta_{i,1}$, which is an influential coefficient, GWPR had small RMSE values when $N \leq 500$. In contrast, GWPR tended to have larger RMSE values than PR when $N = 200$; this tendency worsened for the less influential coefficient $\beta_{i,2}$. Conventional GWPR is recommended when the sample size is sufficiently large (e.g., $N \geq 500$).

The L-GWPR specifications have at least similar RMSE ranges to PR, even when $N = 200$, and are more stable than conventional GWPR. However, the RMSEs of many L-GWPR specifications were larger than those of the conventional GWPR when $N = 500$ and 2,000. Exceptionally, L-GWPRR and L-GWPRR_loc exhibited smaller RMSE values than GWPR in all cases. The accuracy of these specifications was also confirmed.

The CC and RMSE values have similar tendency under the "many zero" scenario ($\mu_{0(\beta)} = -1$) as explained in Supplementary Material 1. L-GWPRR and L-GWPRR_loc had slightly higher CC values and smaller RMSE values than GWPR.

Figure 5 compares the standard deviation (SD) of the coefficients estimated from GWPR and L-GWPRR with the true SD. The conventional GWPR tended to have larger SD values than the true SD values, and variance inflation was prominent for small samples. Therefore, the GWPR coefficients should be interpreted cautiously when the sample size is small. Conversely, the L-GWPRR coefficients tended to have smaller SD values because of the regularization. However, the SD value for L-GWPRR was closer to the true SD value and more stable than that of the conventional GWPR. These results suggest that L-GWPRR successfully stabilized the GWPR estimator through regularization.

*4.2.2: Bandwidth*

Figure 6 displays boxplots of the bandwidths estimated using GWPR and L-GWPRR. The bandwidth estimates are not necessarily close to the range parameter *r*, which is used for data generation. This is because the bandwidth parameter, which determines the sample weights, and the range parameter, which determines the spatially dependent structure, are two different parameters. Both models tended to have similarly small bandwidth values. For GWPR and L-GWPRR, the medians of the optimal bandwidths were {0.05, 0.13} for $r = 0.5$, {0.10, 0.14} when $r = 1$, and {0.10, 0.16}. The slightly larger bandwidth of the L-GWPRR may be more reliable based on the better estimation accuracy of the coefficients.

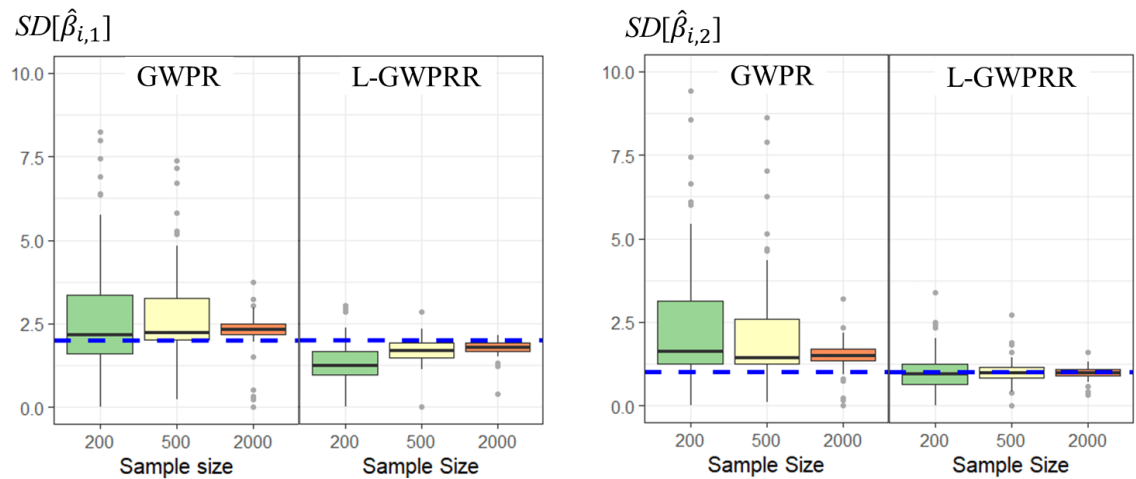

Figure 5. SDs of the estimated coefficients when $\mu_{0(\beta)} = -1$ and $b = 1.0$. The SD values are evaluated in each iteration and summarized in the boxplots. The blue dashed line represents the true SD values.

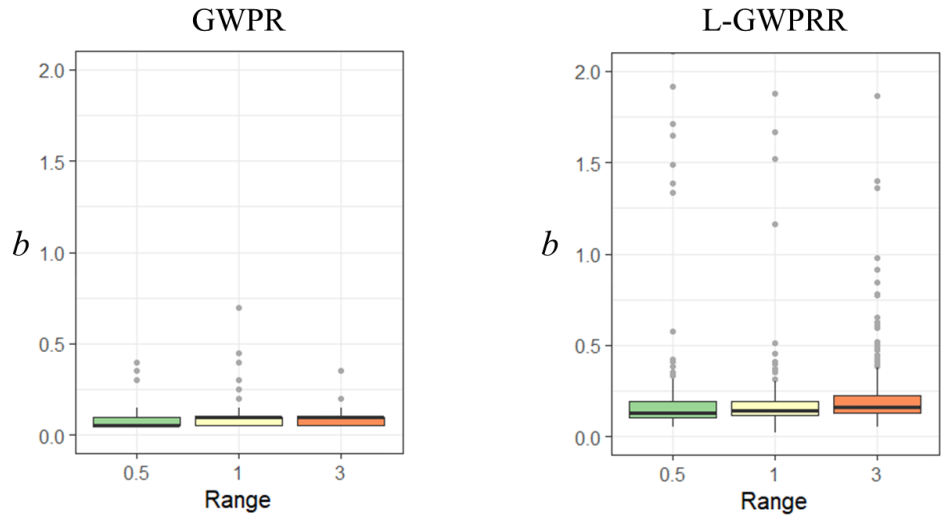

Figure 6: Boxplots of the estimated bandwidth ($\mu_{0(\beta)} = 2$ and $N = 500$).

*4.2.3: Computation time*

Finally, Figure 7 compares the computing times of GWPR, L-GWPR, and L-GWPRR, assuming $N \in \{50, 200, 500, 1{,}000, 2{,}000, 4{,}000, 8{,}000\}$. To verify the computational improvement by replacing the usual double-loop LOOCV with the single-loop LOOCV, we calculated the computation time of the GWPR with LOOCV-based bandwidth optimization implemented in the GWmodel package (Lu et al., 2014; Gollini et al. 2015), which is widely used to estimate the GWPR, and we consider it a reasonable benchmark.[4] A MacBook Pro (3.3 GHz, Intel Core i7, 16 GB memory) was used for this comparison. As expected, the L-GWPR was much faster than the GWPR. Although the L-GWPRR must optimize the ridge and bandwidth parameters, it is faster than the GWPR, which optimizes only the bandwidth. Notably, L-GWPRR is easily accelerated by replacing the linear GWR model (Eq. 13), which is for LOOCV, using the fast GWR model (Li et al., 2019) or scalable GWR model (Murakami et al., 2021).

---

[4] The computation time of GWR4 strongly depends on the range and spacing of the grid in the grid search, while our method does not assume such a grid. GWR4 is not suitable for computation time comparison. Instead, we used GWmodel, which numerically optimizes the bandwidth parameter as with our method.

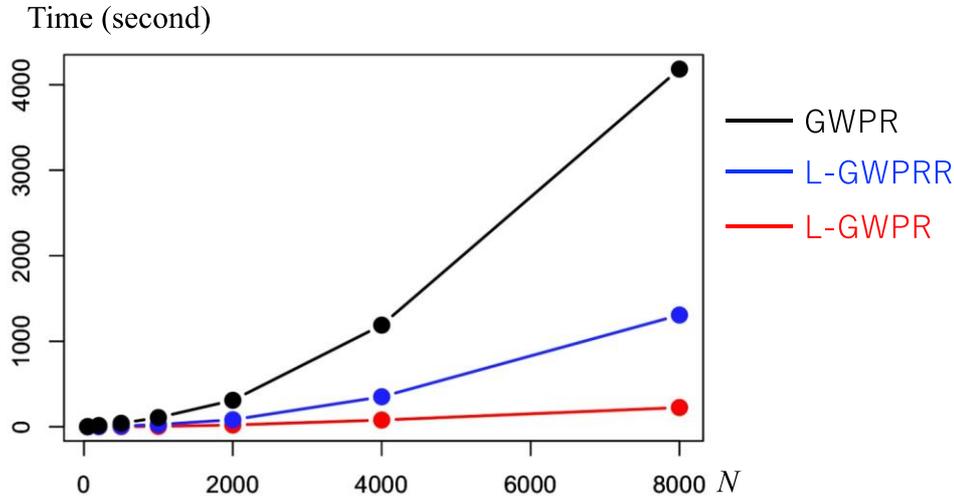

Figure 7. Computing time comparison of GWPR, L-GWPRR, and L-GWPR for coefficients estimation, standard error evaluation, and likelihood evaluation. GWPR is embedded in C++ via the Rcpp package and implemented in the GWmodel package, whereas ours are coded purely by R.

*4.2.4: Summary*

In summary, this section compares L-GWPR and its variants with PR and GWPR. The results showed that L-GWPRR (regularized L-GWPR) tends to be faster and more stable than conventional GWPR. Supplementary Material 2 provides a summary of the results of additional Monte Carlo experiments evaluating the coefficient estimation accuracy of the L-GWPR models using a bisquare instead of the Gaussian kernel specified here (where bisquare and Gaussian kernels are considered the most typically specified in practice). The data-generating process used in Eq. (20) was assumed. The results confirm that the L-GWPRR is the most accurate of all the study models. Therefore, the study results are considered robust to the kernel function choice. The next section empirically compares the GWPR and L-GWPRR specifications.

## 5: Application to a crime analysis study

*5.1: Outline*

This section employs conventional PR, GWPR, and L-GWPRR to an analysis of the number of shopliftings for 1,447 districts, which are called Cho, in the Tokyo Metropolis in April 2019, Japan (Source: the Dai-Tokyo Bouhan network database (https://www.bouhan.metro.tokyo.lg.jp/). Figure 8 shows the number of shopliftings in the target month, where the eastern area surrounded by the Yamanote line, which is a looping railway, is the central area. The analysis did not include Western Tama, a mountainous area where the number of crimes is limited.

The number of trials was used as an offset variable in the analysis. The resulting regressions quantify the influence of each explanatory variable on the number of shopliftings per retail store, and the influences are expected to vary spatially for GWPR and L-GWPRR. The explanatory variables were as follows: number of shopliftings in the previous month (Repeat), nighttime population density (Npden [1,000 people/km$^2$]), daytime population density (Dpden [1,000 people/km$^2$]), and the unemployment ratio (Unemp). Crime tends to be repeated in the same neighborhood, which is known as repeat and near-repeat victimization (Townsley et al., 2003). Hino and Amemiya (2019) verified these repetitive tendencies in residential burglaries in Japan, whereas Amemiya et al. (2020) verified sexual crimes. The estimated coefficients of Repeat attempts quantify such repetitive tendencies. The coefficients of Npden, Dpden, and Unemp quantify the influence of people's density and economic status, which are potentially associated with the risk of shoplifting. These variables are likely to be collinear at some scale (i.e., the likely value of the L-GWPRR implementation). The data were collected from the 2015 National Census Statistics.

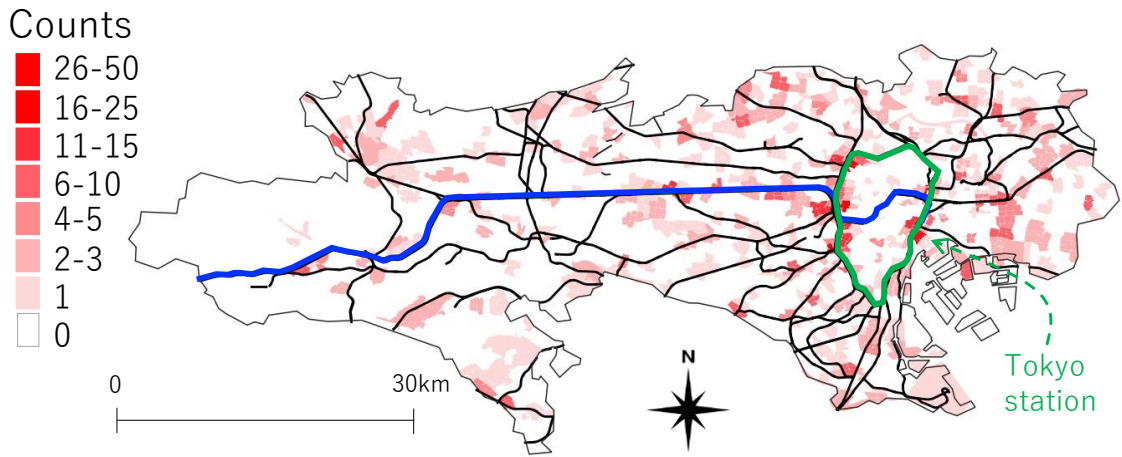

Figure 8. Number of shopliftings by district. The counts are zeros in 1,052/1,447 districts. Japan Railways lines are shown in black lines. Other private railways and subways are not indicated because their networks are too dense to show. The green line denotes the circular Yamanote line, and the blue line denotes the Chuo line. Tokyo station is the central station. There are major commercial and business areas inside or along the Yamanote line, while residential areas are outside the line.

*5.2: Results*

GWPR took 209.9 seconds for the modeling, whereas L-GWPRR took only 52.7 seconds despite optimizing an additional ridge parameter. Therefore, the computational efficiency of the proposed method was confirmed.

     Table 1 lists the estimated parameters for PR, GWPR, and L-GWPRR. For these models, most coefficients of Repeat and Dpden are positive, suggesting that shoplifting tends to be repeated in the same district and that more shoplifting occurs in populated districts. The maximum or minimum values of the GWPR coefficients are often extremely large or small, whereas those of the L-GWPRR coefficients are relatively small in absolute terms. This suggests greater stability of the L-GWPRR coefficients. The optimized bandwidth values are 2.17 km for L-GWPRR and 2.02 km for GWPR,

indicating small-scale map patterns that are estimated in the coefficients for both models.

Regarding model accuracy, both deviance and pseudo $R^2$ indicated that L-GWPRR had the highest accuracy compared with PR and GWPR. The dispersion parameter of L-GWPRR was 0.79, implying under-dispersion (mean > variance); based on the accuracy, this is because the L-GWPRR mean function explains most of the data variation, and the resulting residual variance is small. In short, the importance of considering spatial heterogeneity in regression coefficients and variance dispersion was confirmed in this study.

Table 3. Summary of the parameters estimated from PR, GWPR, and L-GWPRR. Q1 and Q4 mean 1st and 4th quantile, respectively. We used the pseudo $R^2$ of McFadden (1973) for accuracy evaluation that is defined as $1 - D(\hat{\lambda}_i)/D_0$, where $D(\hat{\lambda}_i)$ is the residual deviance (Eq. 17) and $D_0$ is the null deviance.

|            | PR      | GWPR    |       |       |      |        | L-GWPRR |       |       |       |       |
|------------|---------|---------|-------|-------|------|--------|---------|-------|-------|-------|-------|
|            |         | Min     | Q1    | Med   | Q3   | Max    | Min     | Q1    | Med   | Q3    | Max   |
| Const.     | -3.74   | -24.35  | -2.35 | -1.31 | 2.67 | 4.95   | -10.6   | -7.17 | -6.09 | -4.82 | -2.16 |
| Repeat     | 0.06    | -1263.5 | 0.15  | 0.28  | 0.43 | 1.65   | -2.65   | 0.14  | 0.26  | 0.46  | 2.03  |
| Npden      | -0.01   | -0.54   | -0.04 | 0.02  | 0.05 | 36.45  | -0.47   | -0.07 | -0.00 | 0.07  | 0.83  |
| Dpden      | 0.00    | -329.05 | 0.00  | 0.01  | 0.03 | 1.30   | -1.12   | 0.00  | 0.01  | 0.05  | 1.45  |
| Unemp      | 0.80    | -210.08 | -8.74 | 1.06  | 1.71 | 261.34 | -17.36  | -1.46 | 0.70  | 3.48  | 31.61 |
| Bandwidth  | N.A.    | 2.02    |       |       |      |        | 2.17    |       |       |       |       |
| Dispersion | N.A.    | N.A.    |       |       |      |        | 0.79    |       |       |       |       |
| Deviance   | 2649.00 | 1535.49 |       |       |      |        | 1267.42 |       |       |       |       |
| Pseudo $R^2$ | 0.42  | 0.79    |       |       |      |        | 0.83    |       |       |       |       |

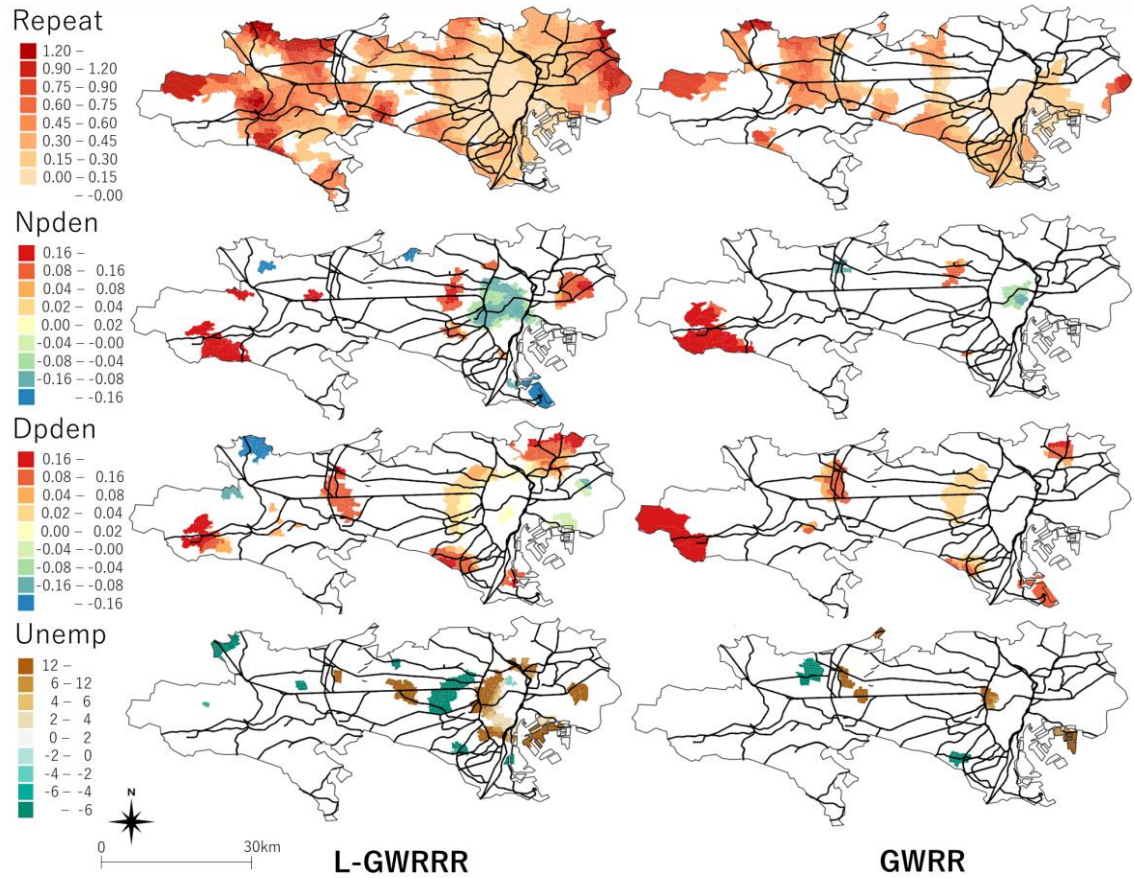

Figure 9. Estimated coefficients. Following Mennis (2006), only those that are statistically significant at the 5 % level are plotted.

Figure 9 (left) shows the statistically significant coefficients estimated using the L-GWPRR model. The statistical significance is evaluated through a z-test, where z-values are evaluated by $z_{i,k} = \hat{\beta}_{i,k}/\sqrt{Var(\hat{\beta}_{i,k})}$ (see Eq.15). To avoid Type I error (i.e., overestimation of statistical significance) due to multiple testing by site and covariate, the p-values were corrected using the procedure proposed by Da Silva and Fotheringham (2015). The coefficients of Repeat are positive and significant in most districts, suggesting that shoplifting tends to be repeated in the same district. Smaller coefficients in the central area of Tokyo suggest that a repetitive tendency is relatively weak, reflecting that the center has many retailers with a reduced criminal need to commit shoplifting in the same district/retail.

The coefficients of Npden and Dpden tend to be positively significant in urbanized residential areas outside the circular Yamanote line (see Figure 8). In these areas, instances of shoplifting tended to increase in local centers with dense daytime and night-time populations. Conversely, in central Tokyo, inside the Yamanote line, the negative coefficients for Npden indicate that shoplifting increases in less-populated areas, where theft can easily go unnoticed. Unemp shows a significantly positive relationship between the unemployment ratio and shoplifting in central commercial areas along the Yamanote line.

Figure 9 (right) shows the coefficients estimated using the conventional GWPR model. The spatial patterns of the coefficients were broadly similar to those for the L-GWPRR, partly because of the similarly small bandwidth estimates (2.17 km for the L-GWPRR and 2.02 km for the GWPR). This similarity in the coefficient maps indicates that L-GWPRR can robustly approximate GWPR in terms of its spatial outputs.

## 6: Concluding remarks

This study developed a computationally efficient and stable alternative to the conventional GWPR by introducing a linear approximation. Monte Carlo experiments demonstrated that our linearized GWPR models (L-GWPR) could accurately estimate local coefficients and that the accuracy is an improvement over conventional GWPR for small samples. Ridge regularization (L-GWPRR) was also observed to considerably improve the coefficient estimation accuracy of linearized GWPR models.

The main advantage of the proposed methodology is its expandability. Our model (Eq. (10)) is a simple linear GWR model that can be easily extended. For example, following Fotheringham et al. (2017), the proposed model can be extended to a multiscale GWPR that estimates the spatial scale (or bandwidth) of individual coefficients (see Comber et al., 2021). A spatiotemporal extension is also possible by meshing our linear model with a geographically and temporally weighted regression model (Huang et al., 2010; Fotheringham et al., 2015). Robust estimations can be incorporated following Harris et al. (2010) and Sugasawa and Murakami (2021).

For large samples, the computational cost can be further reduced by following Li et al. (2019) and Murakami et al. (2020), who proposed fast and scalable GWR, respectively. In general, local models are known to be non-degenerative, meaning that they tend not to suffer from over-smoothing because they ignore local heterogeneity and can accurately estimate local features even from very large samples (see Liu et al., 2020a). An L-GWPR/L-GWPRR extension for large samples, such as one million, is an important future direction.

However, the issue of GWPR stability should be further investigated. Although Fotheringham and Oshan (2016) demonstrated the robustness of GWR against multicollinearity among explanatory variables, others did not find this to be the case (e.g., Bárcena et al., 2014). Murakami and Griffith (2021) suggested that GWR can be unstable when the explanatory variables are spatially dependent. Harris (2019) provides further evidence in this respect. Therefore, in the future, the robustness of GWPR and L-GWPR/L-GWPRR against this problem has to be investigated in more detail.

Finally, L-GWPR/L-GWPRR applies to a wide variety of spatial processes for count data to investigate the regression relationship heterogeneity. For example, an application to data from the COVID-19 pandemic that analyzes the influence of local or regional properties, such as people flow and vaccination ratio, on the number of cases or deaths would be worthy (see Liu et al., 2020b; Zhang et al., 2021; Xu et al. 2021). Such an analysis is useful when considering local rather than global countermeasures.

**Data availability statement**

The R code that implements the proposed models and the data used in Section 5 are available from https://figshare.com/s/991847c3aeb6d7022da4.

**Disclosure Statement**

The authors declare no conflicts of interest associated with this manuscript.

**Supplementary Material 1: Monte Carlo experiments under "many zero" scenario**

Figures A1 and A2 summarize the CC and RMSE values under the many zero scenario ($\mu_{0(\beta)} = -1$). As with the "few zero" scenario, GWPR has reasonably high CC values, whereas L-GWPRR and L-GWPRR_loc have slightly higher CC values. The RMSE values for GWPR showed large variations relative to the L-GWPRR specifications, which could be attributed to the identification problem. Again, L-GWPRR and L-GWPRR_loc had the lowest median RMSE values across all cases.

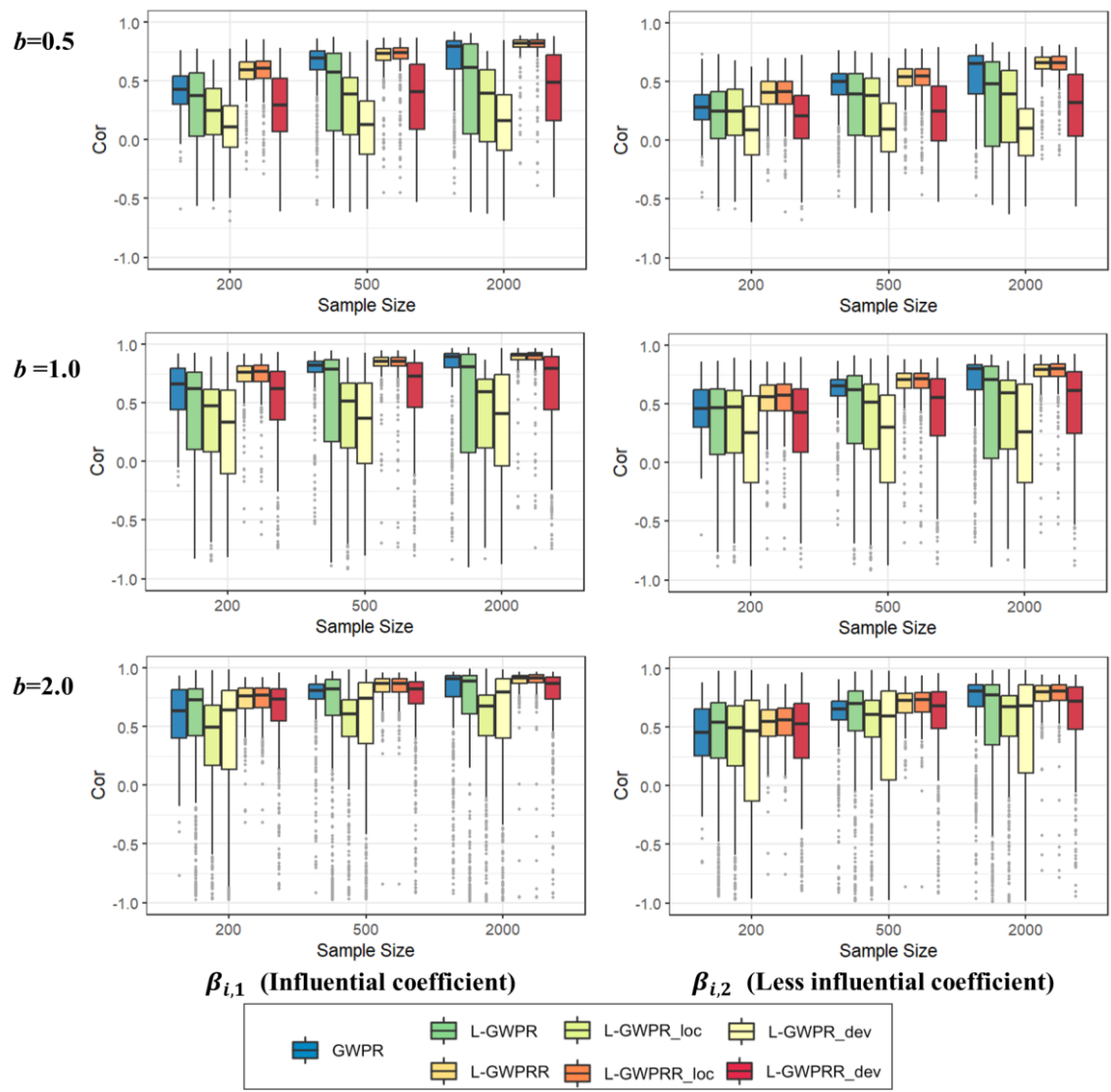

Figure A1. Boxplots of the CCs between $\hat{\beta}_{i,k}$ and $\beta_{i,k}$ in each iteration when $\mu_{0(\beta)} = -1$ (Left: influential coefficient ($\beta_{i,1}$); right: less influential coefficient ($\beta_{i,2}$)).

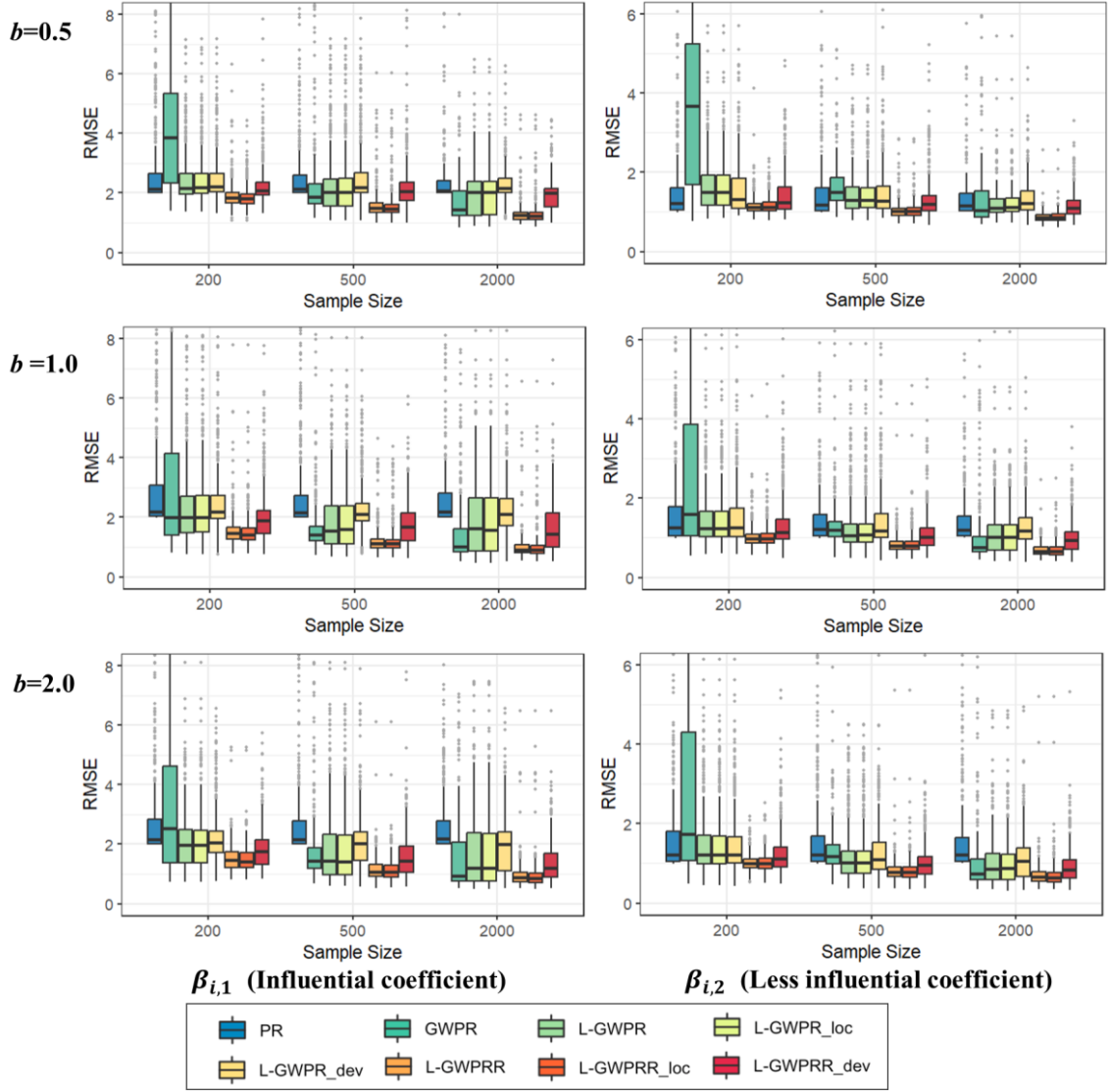

Figure A2. Boxplots of RMSE in each iteration when $\mu_{0(\beta)} = -1$ (Left: influential coefficient ($\beta_{i,1}$); right: less influential coefficient ($\beta_{i,2}$)).

**Supplementary Material 2: Monte Carlo experiments when a bisquare kernel is assumed**

This section presents additional Monte Carlo experiments to evaluate the estimation accuracy of GWPR and L-GWPR using the bisquare kernel instead of the Gaussian kernel. The bisquare kernel is defined as follows:

$$w_{i,j}(b) = \begin{cases} \left(1 - \frac{d_{i,j}}{b}\right)^2 & if\ d_{i,j} < b \\ 0 & if\ d_{i,j} \geq b \end{cases}, \qquad (A1)$$

The data-generating process used in Eq. (19) was assumed. As in Section 4, the models were fitted 200 times for each scenario assumed in Section 4. Figures A3 and A4 plot the mean CC and RMSE values of the influential coefficient $\beta_{i,1}$, and Figures A5 and A6 summarize those for the less influential coefficient $\beta_{i,2}$. As in the case assuming a Gaussian kernel, L-GWPRR and L-GWPRR_loc indicated high CC and low RMSE values relative to the alternatives across all cases. These results confirm the accuracy of the L-GWPRR specifications.

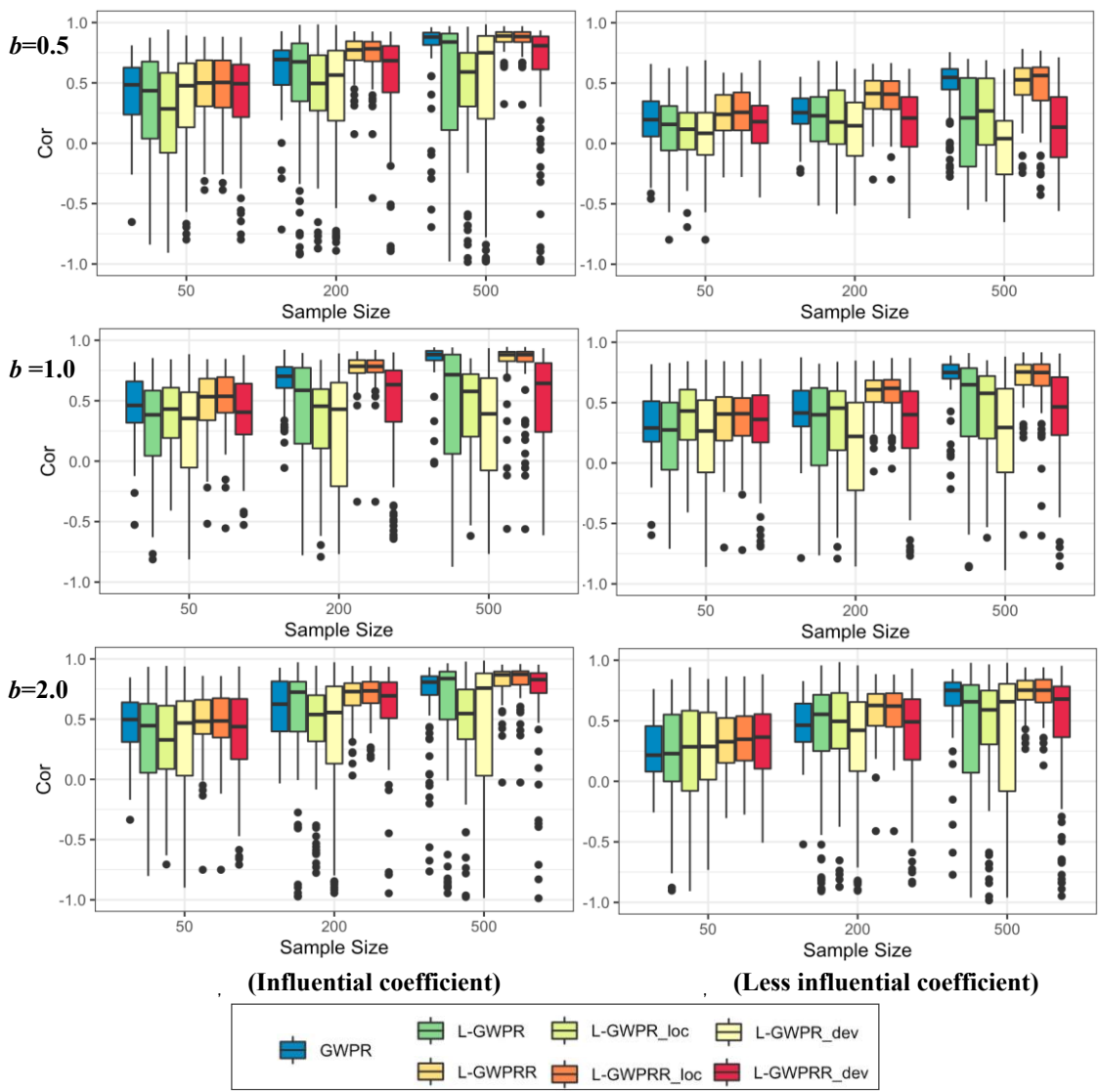

Figure A3. Boxplots of the CCs between $\hat{\beta}_{i,k}$ and $\beta_{i,k}$ when the bisquare kernel is used ($\mu_{0(\beta)} = 2$).

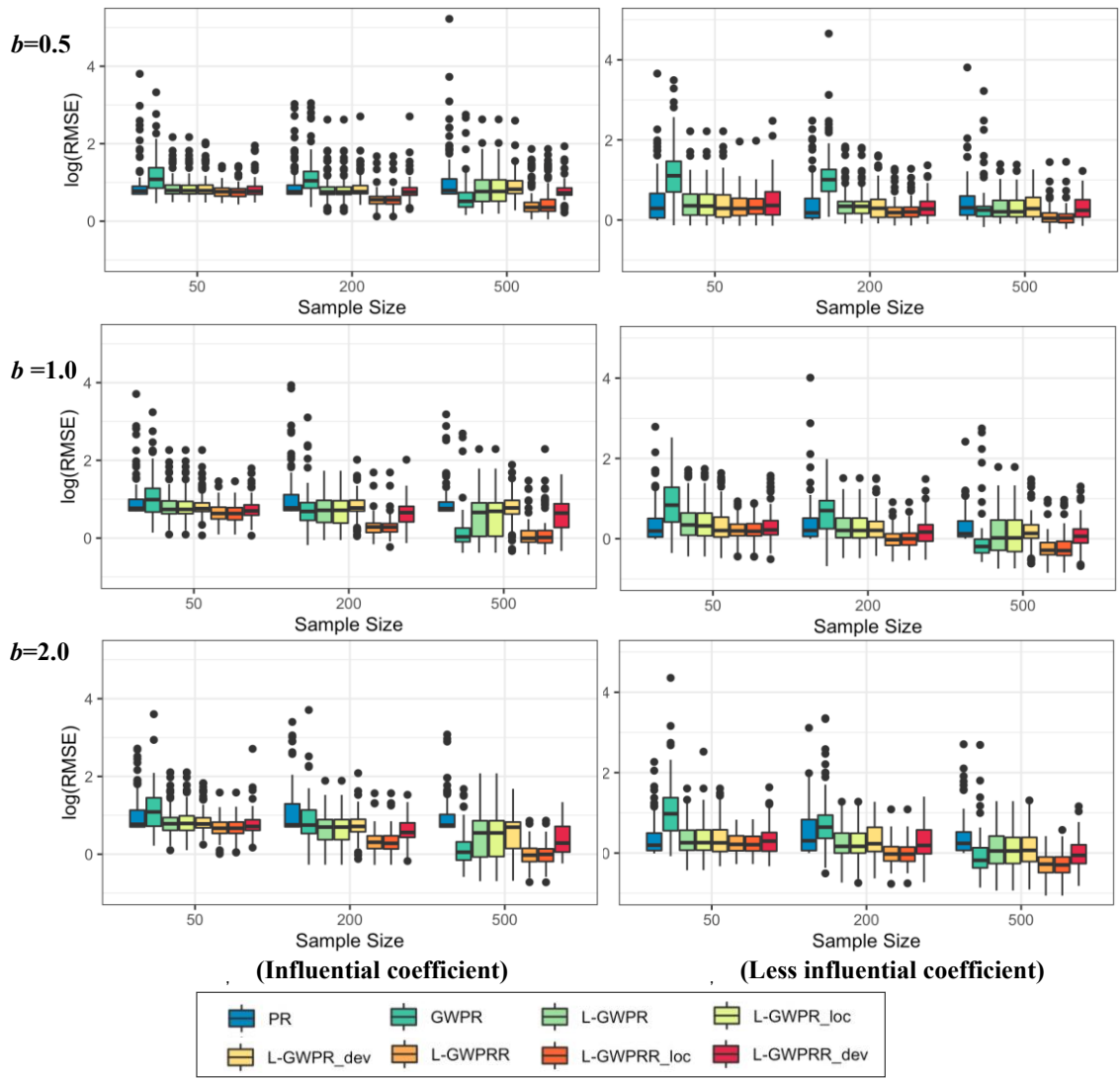

Figure A4. Boxplots of log(RMSE) when the bisquare kernel is used ($\mu_{0(\beta)} = 2$).

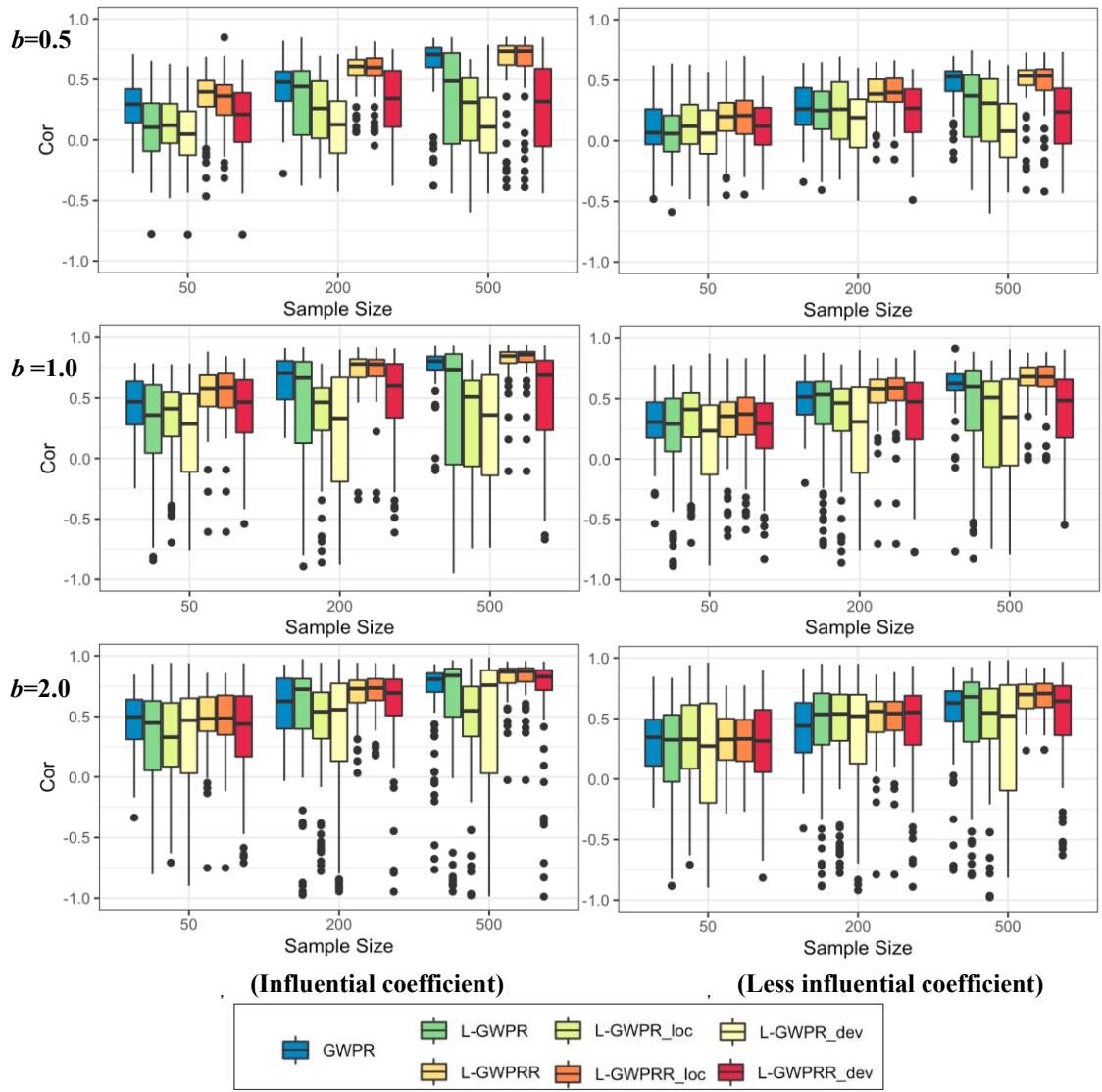

Figure A5. Boxplots of the CCs between $\hat{\beta}_{i,k}$ and $\beta_{i,k}$ when the bisquare kernel is used $(\mu_{0(\beta)} = -1)$.

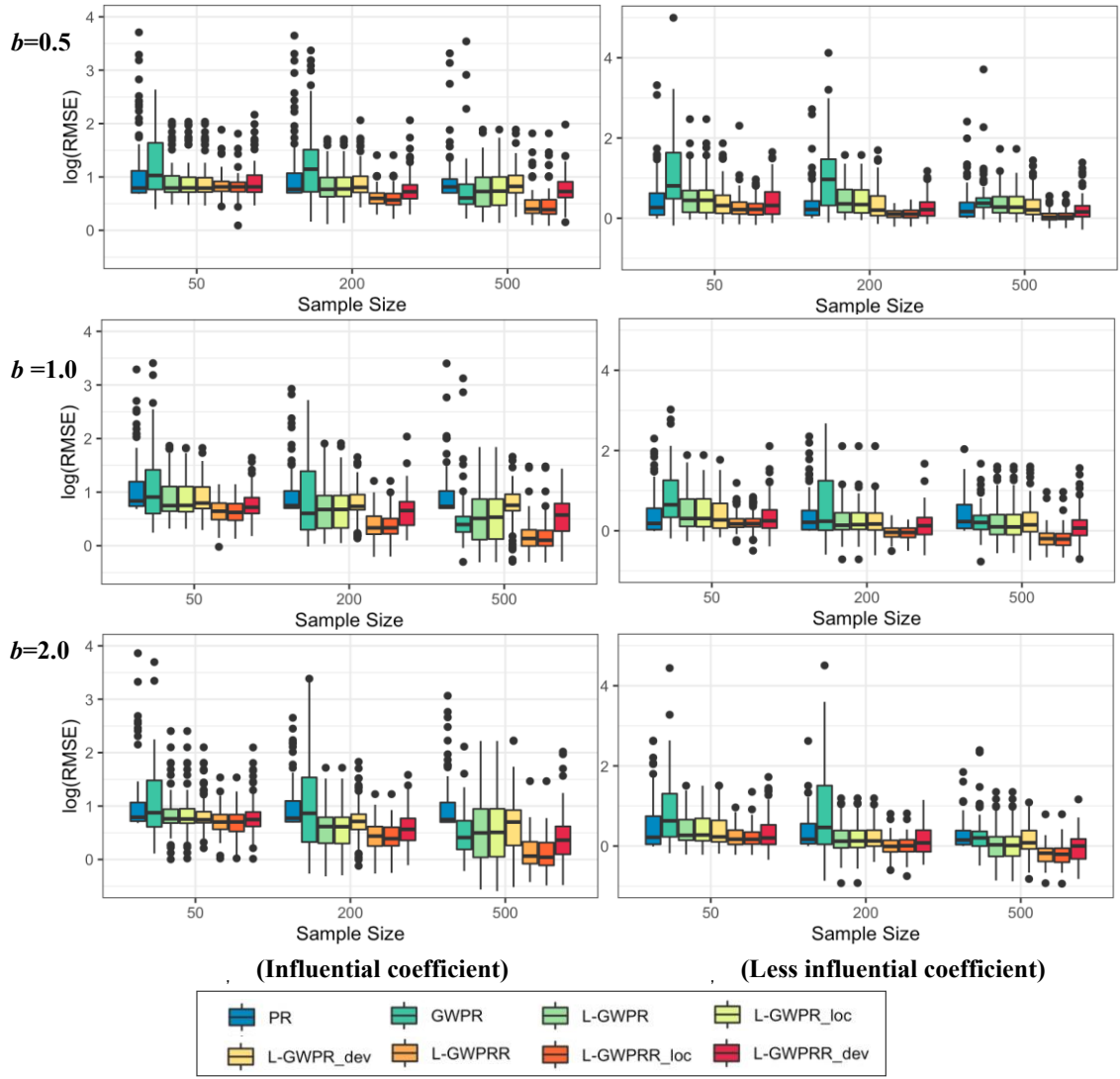

Figure A6. Boxplots of log(RMSE) when the bisquare kernel is used ($\mu_{0(\beta)} = -1$).